\newcommand{\za}{\alpha}

\newcommand{\zd}{\delta}
\newcommand{\zs}{\sigma}

\newcommand{\ze}{\epsilon}

\newcommand{\zt}{\tau}

\newcommand{\zN}{I\hskip-3.4pt N}
\newcommand{\zR}{I\hskip-3.4pt R}

\documentclass[10pt]{article}
\usepackage{epsf}
\usepackage{epsfig}
\title{Gibbs Entropy and Irreversible Thermodynamics}
\author{L. Rondoni \\ Dipartimento di Matematica, Politecnico di Torino\\
Corso Duca degli Abruzzi 24, I-10129 Torino, Italy \\ \\
e-mail:  rondoni @ polito.it \\ \\ 
E.G.D. Cohen \\
Rockefeller University, New York, New York 10021 - U.S.A. \\ \\ \\
AMS numbers: 82C05, 80A20, 70F25 \\
Physics abstracts numbers: 05.45.+b, 05.60.+w, 05.70.Ln}

\begin{document}
\maketitle
\baselineskip=1\baselineskip

\centerline{\small {\bf Abstract}}

{\small Recently a number of approaches has been developed to connect the
microscopic dynamics of particle systems to the macroscopic properties 
of systems in nonequilibrium stationary states, via the theory of
dynamical systems. This way a direct connection between dynamics and 
Irreversible Thermodynamics has been claimed to have been found.
However, the main quantity used in these studies is a 
(coarse-grained) Gibbs entropy, which to us does not seem 
suitable, in its present form, to characterize nonequilibrium states.
Various simplified models have also been devised to give explicit examples
of how the coarse-grained approach may succeed in giving a full 
description of the Irreversible Thermodynamics.
We analyze some of these models pointing out a number of difficulties
which, in our opinion, need to be overcome in order to establish a physically
relevant connection between these models and Irreversible Thermodynamics.}

\vskip 25pt

\section{Introduction}
In recent years, important connections have been made between the 
theory of chaotic dynamical systems and the statistical mechanics
of systems in nonequilibrium stationary states. This is based on 
the widely accepted belief that the dynamics of the microscopic
constituents of matter is chaotic, as also formally expressed by 
the following \cite{GC}:

\noindent
{\bf Chaotic Hypothesis (Gallavotti-Cohen, 1995):}
{\it A reversible $N$-particle system
in a stationary state can be regarded as a transitive Anosov
system, for the calculation of its macroscopic properties.}

Although the dynamical systems methods have led to many interesting 
insights of physical interest,
their application to elucidate the behavior of macroscopic systems,
as done in statistical mechanics or (Irreversible) Thermodynamics
has lead to difficulties which, in our opinion, have not yet been 
fully resolved. There seems to be, then, a qualitative difference 
between pure dynamics and thermodynamics (see, e.g.\ \cite{CR} 
for some facet of this difference not considered here).

In this paper, we will try to clarify some aspects of the recently
developed attempts to incorporate Irreversible Thermodynamics (IT) into
the framework of dynamical systems theory. In this connection we will
concentrate on the interesting recent works by Gaspard (G)
\cite{PGpap,PG}; by Breymann, T\'el and Vollmer (BTV) \cite{VTB,BTV}; 
and especially by Gilbert and Dorfman (GD) \cite{GD}, who 
extensively investigated the connection between a coarse-grained
``entropy'' and IT
in nonequilibrium states. Early works relevant to our discussion
had already appeared in 1996, cf.\ \cite{BTV96}.

The concept of coarse-grained entropy in the study of 
nonequilibrium systems has been discussed in the past.
As a matter of fact, Gibbs himself introduced a coarse grained 
entropy to circumvent the difficulty that, the Gibbs
entropy $S_G$ (cf.\ eq.(\ref{SGfine}) below), does not change during
the time evolution of a Hamiltonian system \cite{Tolman}. 
Similarly, the final goal of introducing the coarse graining
in \cite{PGpap}--\cite{BTV96} could be stated as that of 
{\it circumventing certain
difficulties which affect the Gibbs entropy of
nonequilibrium systems, thus
building a complete description of all quantities occurring in IT}
(cf.\ Eq.(\ref{deGMbala}) below) {\it in purely dynamical terms.}
The guiding idea in this endeavor is the identification of the 
irreversible entropy
production rate with a special form of loss of information rate, to be 
defined below (cf.\ subsection 3.1).
We begin our analysis with a description of the results obtained so far
with the coarse grained approach, and then consider
the difficulties which we find with it. This way, we indicate what might
have to be considered further, in order to obtain a consistent theory of IT.

We note that a coarse-grained
description --both in space and time-- is also at the basis of IT
itself \cite{DEGM}. Indeed,
the basic equation for the entropy change in IT is [10(a)]
\begin{equation}
\frac{\Delta_{tot} S}{\tau} = \frac{1}{\tau} \left[
\Delta_e S + \Delta_i S \right] ~ ,
\label{deGMbala}
\end{equation}
where we have divided by a small but finite time
$\zt$ to obtain the rate of entropy change. Here, $\Delta_e S$ is the 
entropy exhanged by the system with its surroundings, while $\Delta_i S$
is, respectively, the entropy produced inside the system, in a time $\zt$.
This relation can be re-written in the more usual local differential form 
as \cite{DEGM}:
\begin{equation}
\frac{\partial \rho s}{\partial t} = - \mbox{div } {\bf J}_{s,tot}
+ \zs ~; \quad \zs \ge 0
\label{DEGMlocal}
\end{equation}
where $\rho$ is the density of the system, $s$ is the entropy per
unit mass, ${\bf J}_{s,tot}$ is the total entropy flow rate per unit
area corresponding to the term $\Delta_e S/\zt$, and $\zs$ is
the entropy production rate per unit volume.
In particular, for a diffusive system, the term
$\zs$ can be related to the gradients in space of the densities of the 
various diffusing substances.
Therefore, space derivatives of various quantities appear
in the expressions for the entropy flow and entropy production rates. 
Equation (\ref{DEGMlocal}) can also be written as
\begin{equation}
\rho \frac{d s}{d t} = - \mbox{div } {\bf J}_{s}
+ \zs ~, \quad {\bf J}_s = {\bf J}_{s,tot} - {\bf J}_{s,c}
\label{DEGMloc2}
\end{equation}
where ${\bf J}_{s,c} = \rho s {\bf v}$ is the convective flow,
and ${\bf v}$ is the fluid velocity.

\section{Gibbs entropy and (nonequilibrium) dynamical systems}
We begin with a dynamical system $(X,\phi^t)$ representing the dynamics
of an $N$-particle system in $3$ dimensions. Then, $X \subset \zR^{6N}$
is the phase space of the system, and $\phi^t$ is an invertible 
transformation of $X$ into itself for all times $t \in \zR$. 
Given a probability measure $\mu_0$ on $X$ at time $0$, the dynamics of 
the system induces an evolution which in terms of the measurable
sets $A \subset X$ can be expressed by
\begin{equation}
\mu_t(A) = \mu_0(\phi^{-t} A) ~, \quad t \in \zR ~.
\label{measpace}
\end{equation}
This expression defines the time evolution of the probability 
distribution in phase space, such that the ``mass'' in the set
$A$ at time $t$, $\mu_t(A)$, is the same as it was in $\phi^{-t} A$ 
at time zero, $\mu_0(\phi^{-t} A)$.
The measure $\mu_t$ can be seen as characterizing the state
of the particle system at time $t$, in the sense that the expectation
values of the ``observables'' ${\cal O}$
of the system (e.g.\ smooth functions
of phase, ${\cal O} : X \rightarrow \zR$) are given as averages of 
such functions with respect to $\mu_t$.
 
\vskip 5pt
\noindent
{\bf Gibbs Entropy: }{\it If $\mu_t$ has a density $\rho_t$ on $X$,
i.e.\ $\mu_t(d x) = \rho_t(x) d x$, the Gibbs entropy of the system at
time $t$ is defined by the quantity 
\begin{equation}
S_G(t) = - k_B \int_X \rho_t(x) [\log \rho_t(x)-1]dx
\label{SGfine}
\end{equation}
where $k_B$ is Boltzmann's constant.}\footnote{The constant ``$-1$'' in
the integrand of Eq.(\ref{SGfine}) is introduced only for consistency
with the definitions of \cite{PGpap,PG,GD}.}
\vskip 5pt
\noindent
We refer to $S_G$ as to a 
{\em fine grained} quantity to emphasize its difference from the coarse 
grained quantities defined below (e.g. Eq.(\ref{Aprime})), 
in the sense that its definition involves an integral over $X$ instead of 
a sum over a partition by finite-volume sets of $X$.

Unfortunately, 
the stationary states of the current models of nonequilibrium physical
systems, seen as dynamical systems, are represented by singular 
measures $\mu$ for which
eq.(\ref{SGfine}) does not make sense \cite{PG,BTV,GD}. It is then 
argued (cf.\ \cite{PG}, Section 8.6) that coarse grained entropies
should be used to characterize these nonequilibrium stationary states
{\it ``... especially if we want to keep the operational interpretation
of entropy as a measure of disorder.''}
In the following subsections, we describe two classes of 
models, showing how singular measures arise.

\subsection{Thermostatted systems}
Consider an $N$-particle system whose equations of motion contain 
the action of an external force field ${\bf F}^e$ and 
compensating (``thermostatting'') terms, which
eliminate the increase of the (dissipative) energy of the system due
to the work performed on the particles by ${\bf F}^e$, so that the 
system will finally reach a nonequilibrium stationary state \cite{EM}. 
The equations of motion of one such system are:
\begin{equation}
\left\{ \begin{array}{lcl}
{\bf{\dot{q}}}_i & = & {\bf{p}}_i/m \nonumber \\
{\bf{\dot{p}}}_i & = & {\bf{F}}^i_i + {\bf{F}}^e_i -\alpha (x){\bf{p}}_i
\end{array} \right. \, \quad
i=1, ..., N \, ; \quad x \equiv (q,p) \in X \subset \zR^{6N} ~,
\label{eqsmot}
\end{equation}
with periodic boundary conditions, so that $X$ can be assumed to be
compact. 
Here $m$ is the mass of the particles;
${\bf{F}}^i_i$ and ${\bf{F}}^e_i$ are the forces on particle $i$ 
due to the other particles in the system and to the external field, 
respectively; $x \equiv (q,p) \equiv ({\bf{q}}_i, {\bf{p}_i})$, 
$i=1, ..., N$, stands for the collection of all the positions and 
momenta of the particles; and 
$\alpha (x){\bf{p}}_i$ represents the effect of the ``thermostat'' 
on the system. The thermostatting function $\alpha(x)$
is obtained from Gauss' principle of minimum constraint \cite{EM}
and is usually
chosen in such a way that either the kinetic or the total energy of
the system remain constant in time.
We refer to such systems as {\em thermostatted systems}.
For constant total energy [isoenergetic (IE) constraint], one obtains:
\begin{equation}
\za(x) = \alpha_{IE}(p) = {\frac{\sum^N_{i=1}\frac{{\bf{p}}_i}{m} 
\cdot {\bf{F}}^e_i} {\sum^N_{i=1} \frac{{\bf{p}}^2_i}{m}}} ~,
\label{alfIE}
\end{equation}
which shows that $\alpha(x)$ is of order $O(1)$ and
is related to the dissipation or the
(generalized) entropy production rate {\em in the system}.\footnote{We speak 
of generalized entropy and of kinetic temperature below
because our systems are
not necessarily close to equilibrium. Our definition of kinetic temperature 
can be modified to involve the peculiar velocities, if 
the center of mass of the system is not at rest \cite{EM}.}
Indeed, if we define  the current (flux in IT) at time $t$ as
${\bf J}_t = \langle \sum_i {\bf{p}}_i/m \rangle_t$ (an average with
respect to the time dependent distribution $\mu_t$) 
and similarly set the average 
$\langle \sum^N_{i=1} \frac{{\bf{p}}i^2}{m} \rangle_t$ equal to
$3N k_BT_t$, where $T_t$ is the {\em kinetic} temperature of the 
system at time $t$,
then, for a constant external field ${\bf{F}}^e$ (force in IT) and for
a large system (large $N$) \cite{CR}, we can write:
\begin{equation}
\langle \za_{IE} \rangle_{t} = 
\frac{\bf{J_t} \cdot \bf{F}_e}{3N k_BT_t} ~,
\label{avealf}
\end{equation}
which yields the IT entropy production rate per degree of
freedom at time $t$.

Starting from a distribution $\mu_0$ on $X$ with density $\rho_0$,
the time evolution of the dissipative system Eqs.(\ref{eqsmot}),(\ref{alfIE})
gradually rearranges the distribution, concentrating it on sets of 
smaller and smaller volume in phase space. This produces what is usually
called a phase space contraction, together with a sequence 
of more and more irregular densities $\{ \rho_t \}_{t > 0}$. In the 
long time limit, a singular distribution is obtained, which assigns
a probability of one to sets of zero phase space volume.
These sets are, in general, dense in $X$
if the external field is not too large, but with decreasing fractal
dimension for increasing fields, till they are not dense anymore 
at high fields (cf.\ \cite{DMR} for the Lorentz gas). 

The rate of variation of $S_G$ for all $t > 0$ is (\cite{EM}, p.252):
\begin{equation}
\dot{S}_G(t) \equiv \frac{d S_G}{d t}(t) = -3 N k_B \langle \alpha
\rangle_t + O(k_B \langle \alpha \rangle_t) ~,
\label{SGrate}
\end{equation}
since the divergence of the equations of motion, Eqs.(\ref{eqsmot}), is 
given by	
\begin{equation}
\mbox{div}~ \dot{x} = -3N \za(x) + O(\za(x)) ~.
\label{diveqm}
\end{equation}
We note that $\dot{S}_G(t)$ converges to a negative constant value,
$\dot{S}_G(t) \approx -3Nk_B \langle \za \rangle_{ss}$ for large $N$  
and large $t$, where the subscript $ss$ in $\langle \za \rangle_{ss}$
indicates the steady state value. 
The result is that $S_G(t)$ diverges to $-\infty$ 
as $t \rightarrow \infty$ \cite{EM}, and it does so in an approximately 
linear fashion after a given relaxation time.
Equation (\ref{SGrate}) is similar in content to Eq.(16) of Goldstein,
Lebowitz and Sinai \cite{GLS} for positive times.

This dynamical description of a system in a nonequilibrium state yields 
the IT expression for the irreversible entropy production rate at any 
instant of time $t > 0$, which is obtained from Eq.(\ref{avealf}). 
Surprisingly,
$\dot{S}_G(t)$ is observed to equal precisely the negative of this 
irreversible entropy production at all times $t$, cf.\ 
Eqs.(\ref{avealf},\ref{SGrate}). Thus, although so far it 
has not been possible to identify a quantity representig the
entropy of the system, a  
connection between IT and an appropriately constructed function, somehow 
related to $S_G$, has been discovered. This, however, is not sufficient
to imply that the entropy of the system should be linked with 
$S_G$. On the contrary, as discussed below in Section 4,
the asymptotic divergence of $S_G$ suggests in fact that attempts to
find such a link are likely to fail.

\subsection{Multibaker maps with flux boundaries}                           
A different class of nonequilibrium models is represented by finite
multibaker chains coupled at both ends to infinite 
``reservoirs'' \cite{PGpap,PG}, i.e.\ chains with 
flux boundary conditions. These models give rise, in the ``macroscopic 
limit'',\footnote{We put in quotes ``reservoirs'' and ``macroscopic 
limit'', as they are crucial for a connection with IT as explained
below.} to stationary states characterized by singular measures in phase 
space,
and are thought to behave similarly, on some respects, 
to certain ideal gas systems, such as the Lorentz gas  
considered by G in Chapter 8 of \cite{PG}.

Several variations of these multibaker systems have been considered. 
We follow G's definitions \cite{PGpap,PG} first. The space
of the multibaker map with flux boundaries $X$ is made of a chain of 
squares $B_n = [0,1] \times [0,1]$, $n \in \{ ... -2,-1,0,1,2, ... \}$
each placed at one site of an infinite one-dimensional lattice, as depicted
in Fig.\ 1. The central section of the chain, whose squares are labelled by
$n=0,1,...,L$, represents a system coupled to two reservoirs: one 
at its left boundary (the squares labelled by $-1,-2,...$), and the other
at its right boundary (the squares labelled by $L+1,L+2,...$). Each $B_n$
contains a certain number of points, thought to represent
noninteracting particles, distributed according to a given distribution 
$\mu(n,x,y)$
defined on $X$, whose time evolution is defined in different
ways for the system and the reservoirs respectively. In practice, one 
time step moves the point $(n,x,y)$ (the point $(x,y)$ of $B_n$) to 
the point $\phi(n,x,y)$, where
\begin{equation}
\phi(n,x,y) = \left\{ \begin{array}{lcr} 
\left( n-1,2x,\frac{y}{2} \right), & 0 \le x < 1/2, & 1 \le n \le L+1 \\
\left( n+1,2x,\frac{y+1}{2} \right), & 1/2 \le x \le 1, & -1 \le n \le L-1 \\
\left( n-1,x,y \right),& 0 \le x < 1/2,& n \le 0 , n \ge L+2 \\
\left( n+1,x,y \right),& 1/2 \le x \le 1,& n \le -2 , n \ge L
\end{array} \right.
\label{MBdynam}
\end{equation}
as depicted in Fig.\ 1.
This dynamics is area preserving.
Starting with appropriate initial point distributions in the infinite
chain, one obtains a system of points coupled to two reservoirs, which 
feed points into the system at the fixed densities, $\rho_+$ (the left
reservoir) and $\rho_-$ (the right reservoir).
By density we
simply mean the number of points per unit area, in each region 
of $X$. During the time evolution a certain density 
profile is created, possibly converging to an invariant distribution
in the long time limit.

Because infinitely many points are required for the
reservoirs,
the measure $\mu$ is not normalized. However, a probability distribution
$\chi$ (cf.\ Eq.(\ref{Poissonsusp}) below)
can still be given for this system, considering the Poisson 
suspension measure associated with 
$\mu$ \cite{PG,CFS}. In this case, the phase space ${\cal M}$ of the system of 
``independent'' points is part of the power set ${\cal P}(X)$ of the 
multibaker space $X$. In ${\cal M}$,
a ``Gibbs entropy'' can be defined as usual, if $\chi$ is not singular.

In the stationary state considered in \cite{PGpap,PG} the density profile 
is made of two kinds of strips only: those having density $\rho_+$ and 
those having density $\rho_-$, which are separated by straight line segments. 
In the squares which are closer to the left reservoir, the strips with 
density $\rho_+$ dominate, while those with density $\rho_-$ dominate
at the other end of
the system, so that $\mu(B_n)$
is linear in the squares' label $n$, for $0 \le n \le L$.
As long as $L$ is
finite, the corresponding Poisson measure $\chi$ is not singular. However, 
in order to obtain results which serve the purpose of nonequilibirum 
statistical mechanics,
singular measures are needed in G's approach (cf.\ \cite{PG} p. 384). 
These are obtained in \cite{PGpap,PG}
through a ``macroscopic limit'', defined by 
$L \rightarrow \infty$ and $(\rho_+ - \rho_-)/L = $constant.
In this limit, the invariant distribution $\mu$ 
becomes singular: the strips with the two different 
densities become thinner and thinner and more and more numerous, while 
$\rho_+$ grows without bounds.
The corresponding Poisson measure $\chi$ is 
also singular, hence $S_G$ cannot be defined.

An interesting generalization of G's model was proposed by BTV \cite{BTV,VTB}. 
The baker space $X$ now consists of a chain of identical rectangles of sides 
$a$, in the horizontal direction and $b$ in the vertical direction,
respectively, Fig.\ 2.
The boundary conditions can still be implemented by two infinite
reservoirs as above. The dynamics are also slightly more general 
(Fig.\ 2). Each rectangle is divided in three vertical
strips of horizontal widths $la$, $sa$ and $ra$ (from left to right),
respectively,
where $l,s,r \ge 0$, and $l+s+r=1$. Each rectangle is also
subdivided into
three horizontal strips of width $a$ and heights $rb$ (bottom strip),
$sb$ (central strip) and $lb$ (top strip).
The leftmost strip of rectangle $m$
is compressed and expanded and moved to fit the bottom horizontal 
strip of  rectangle $m-1$ (nearest left neighbour);
the central vertical strip of rectangle $m$ remains in rectangle $m$,
but is stretched and compressed so that it fits in the central horizontal 
strip; the rightmost vertical strip of rectangle 
$m$ is stretched and compressed to fit the top horizontal strip of
rectangle $m+1$ (right nearest neighbour).
The same procedure is applied to each rectangle of the system, while
the points of the reservoirs are merely translated to the left and the
right,
without volume compression or expansion, like in \cite{PGpap}.\footnote{G's
original model \cite{PG} then corresponds to taking $l=r=1/2$,
and $s=0$.}

Accordingly, after one time step, the distribution has changed
in the chain, and with that the density of points
in each rectangle as well as in each strip has changed. Let 
$\varrho_m$ be the density in rectangle $m$. This density evolves
in the system like
\begin{equation}
\varrho_m(t+\zt) = (1-r-l)\varrho_m(t) + r \varrho_{m-1}(t) -
l \varrho_{m+1}(t) ~.
\label{BTVdens}
\end{equation}
Again, the invariant distribution $\mu$ in the baker space $X$ and the
associated Poisson measure $\chi$, are singular and $S_G$ is not defined. 
However, the mechanism through which the singularities 
are created is not by taking a macroscopic limit like in \cite{PGpap,PG},
but by a combination of phase 
space contraction and boundary effects.

\section{The coarse-grained approach}
To avoid the fact that $S_G$ is not defined in the current models
of nonequilibrium stationary states, as
discussed in Subsections 2.1 and 2.2, several attempts have been
made to replace $S_G$ by a coarse grained {\em information}
entropy\footnote{To clearly distinguish between the physical entropy of IT 
and dimensionless information related entropies, we follow
Nicolis and Daems \cite{ND}, and choose to call information 
entropy a dimensionless entropy-like quantity.}
which takes
finite values in the case of both non-singular {\em and} singular 
distributions \cite{PG,BTV,GD}. This approach is invoked in order 
a) to give a precise meaning to the concept of nonequilibrium entropy 
\cite{BTV}; b) to properly handle the singularities of the stationary 
states, without giving up the interpretation of entropy as a measure 
of disorder \cite{PG} p.370; c) to have a microscopic definition of 
the entropy production rate which agrees with that 
[Eqs.(\ref{deGMbala})-(\ref{DEGMloc2})] of IT \cite{GD}.
This would amend the restriction
encountered in the usual thermostatted systems approach, where
only the irreversible entropy production appears and not the complete
description of IT, as given in Eq.(\ref{deGMbala}). Here,
we will follow GD's approach and notation, which generalizes to some 
extent the previous ones, and emends some aspects of
the original definitions of \cite{PGpap,PG}. 

GD first consider a generating partition, ${\cal A}$, 
for the phase space $X$. Then a discretization of the 
time evolution by time steps of length $\tau$, is introduced to produce 
finer and finer partitions ${\cal A}_{\ell,k}$:
\begin{equation}
{\cal A}_{\ell,k} = \phi^{-l\tau}( {\cal A} ) \vee 
\phi^{-(l-1)\tau}( {\cal A} ) \vee ... \vee {\cal A} \vee ...
\phi^{(k-1)\tau}( {\cal A} )
\label{partit}
\end{equation}
by taking the intersections of the cells of ${\cal A}$ evolved by the 
dynamics of $\phi^\tau$ up to $k-1$ 
time steps forward in time and up to $\ell$ time steps backwards in
time.\footnote{These partitions are intended to be rigid frames 
into which the phase space $X$ is subdivided once and for all. 
They are therefore 
not affected by the time evolution of the system, although the dynamics 
has been used to construct the partitions. Thus, once the partition 
${\cal A}_{\ell,k}$ has been made, it remains in place without 
any change, independently of the dynamical evolution of the system
which takes place ``through it''.}
The symbol
$\vee$ indicates the intersection of all the sets of a given partition 
with those of another one. In particular, we have 
${\cal A}_{\ell+1,k} = \phi^{-\tau} {\cal A}_{\ell,k} 
\vee {\cal A}_{\ell,k}$. Also, GD indicate by $\mu_t$ the phase space
distribution and by $\nu$ the Liouville measure.

\vskip 5pt
\noindent
{\bf GD information entropy: } {\it Consider all the 
sets of the form $B = \cup_i E_i$, with $E_i \in {\cal A}_{\ell,k}$, i.e.\ 
all the sets which are unions of the cells of ${\cal A}_{\ell,k}$. 
On these sets the GD {\em coarse-grained} information 
entropy $S^{GD}_{\ell,k}(B,t)$ is defined by
\begin{equation}
S^{GD}_{\ell,k}(B,t) = - \sum_{A \in {\cal A}_{\ell+1,k} \cap B} 
\mu_t(A) \left[\log \frac{\mu_t(A)}{\nu(A)}- 1\right] ~. 
\label{Aprime}
\end{equation}
where the sum is carried out over all $A \in {\cal A}_{\ell+1,k}$
whose union is $B$.
}

\vskip 5pt
\noindent 
The relation between $S^{GD}_{\ell,k}$ and $S_G$, 
in the case that $\mu_t$ has a density $\rho_t$, is then given by:
\begin{equation}
S_G(t) \equiv k_B S_I(t)
= k_B \lim_{\ell,k \rightarrow \infty} S^{GD}_{\ell,k}(X,t) ~,
\label{SGlim}
\end{equation}
where we also defined the {\em fine grained} information entropy
$S_I$. Hence, for regular measures, the coarse grained 
entropies approximate better and better their fine-grained counterparts, 
when the graining of phase space is made finer and finer. 
On the contrary, if $\mu_t$ is singular, $S_G$ and $S_I$ do not 
exist, while $S^{GD}_{\ell,k}$ for any $\ell,k \in \zN$ does. 
 
The total rate of information entropy
change in a time $\zt$ is then defined by:
\begin{eqnarray}
\frac{\Delta_{tot} S^{GD}_{\ell,k}(B,t)}{\tau} & = & \frac{1}{\tau}
\left[ S^{GD}_{\ell,k}(B,t+\tau) 
- S^{GD}_{\ell,k}(B,t) \right] \nonumber \\
& = & -\sum_{A \in {\cal A}_{\ell+1,k} \cap B} \left[
\frac{\mu_t(\phi^{-\tau}A)}{\tau} \log 
\frac{\mu_t(\phi^{-\tau}A)}{\nu(A)} 
- \frac{\mu_t(A)}{\tau} \log \frac{\mu_t(A)}{\nu(A)} \right]
\label{GDtot}
\end{eqnarray}
where one has used Eq.(\ref{measpace}),
$\mu_{t + \tau}(A) = \mu_t(\phi^{-\tau}A)$, to get the second equality.
This rate of change is then decomposed by GD into a sum of three terms:
\begin{equation}
\frac{\Delta_{tot}S^{GD}_{\ell,k}}{\tau} = \frac{1}{\tau} \left[
\Delta_e S^{GD}_{\ell,k} + \Delta_{th} S^{GD}_{\ell,k} + 
\Delta_i S^{GD}_{\ell,k} \right] ~,
\label{GDbala}
\end{equation}
where, $\Delta_e S^{GD}_{\ell,k}(B)$ is called the change in information 
entropy due to the flow between $B$ and its environment,
$\Delta_{th} S^{GD}_{\ell,k}(B)$, the change in information entropy
due to a thermostat in contact with the system, and
$\Delta_i S^{GD}_{\ell,k}(B)$, that due to irreversible information entropy
production in the system. This separation is based on an interpretation of 
thermostatted equations of motion for particle systems such as 
Eqs.(\ref{eqsmot}), where the thermostatting term
is seen as representing a real thermostat. We remark that $S^{GD}_{\ell,k}$
is defined in terms of the phase space distribution $\mu_t$, hence changes
of this distribution in phase space imply changes in $S^{GD}_{\ell,k}$.

In particular, the information entropy change rate due to flow 
is defined by GD as originally done by G \cite{PG} by:
\begin{eqnarray}
\frac{\Delta_e S^{GD}_{\ell,k}(B,t)}{\tau} & = & \frac{1}{\tau} \left[
S^{GD}_{\ell,k}(\phi^{-\tau}B,t) - 
S^{GD}_{\ell,k}(B,t) \right] \nonumber \\ 
& = & -\sum_{A \in {\cal A}_{\ell+1,k} \cap B} \left[
\frac{\mu_t(\phi^{- \tau}A)}{\tau}
\log \frac{\mu_t(\phi^{- \tau}A)}{\nu(\phi^{- \tau}A)} -
\frac{\mu_t(A)}{\tau} \log \frac{\mu_t(A)}{\nu(A)} \right] ~,
\label{GDflow}
\end{eqnarray}
while the change due to the thermostat is defined by:  
\begin{eqnarray}
\frac{\Delta_{th}S^{GD}_{\ell,k}(B,t)}{\tau} & = & \frac{1}{\tau}
\left[ S^{GD}_{\ell,k}(B,t+\tau) - 
S^{GD}_{\ell+1,k-1}(\phi^{- \tau}B,t) \right] \nonumber \\
& = &  - \frac{1}{\tau} \sum_{A \epsilon {\cal A}_{\ell+1,k} \cap B} 
\mu_t(\phi^{- \tau} A) \log \frac{\nu(\phi^{- \tau}(A))}{\nu(A)} ~.
\label{GDth}
\end{eqnarray}
In the first equality of Eq.(\ref{GDth}), 
the partition ${\cal A}_{\ell,k}$ is compared
with its preimage under $\phi^{-\tau}$, i.e.\ with
${\cal A}_{\ell +1, k - 1} = \phi^{- \tau}{\cal A}_{\ell,k}$, which should
correspond to a different degree of resolution of the phase space.
The term $\Delta_i S^{GD}_{\ell,k}(B,t)/\tau$ in Eq.(\ref{GDbala})
is then deduced from Eq.(\ref{GDbala}) itself, once the other terms
have been defined by the Eqs.(\ref{GDtot}),(\ref{GDflow}),(\ref{GDth}).

\subsection{Gilbert--Dorfman results}
We first discuss the connection of GD's theory with IT. For that, the
term $\Delta_{th} S^{GD}_{\ell,k}(B,t)/\tau$ is crucial. 
Consider thereto the case in which $B=X$,
and the system is in a stationary state characterized by the
natural (invariant) measure $\mu$. If we denote by 
${\cal J}$ the Jacobian determinant
of the transformation $\phi^\zt$, we can write
\begin{equation}
\nu(\phi^{- \tau}(A)) = \int_{\phi^{- \tau}(A)} d x = 
\frac{\nu(A)}{{\cal J}(\phi^{- \tau}(x_{_A}))}
\label{Jacob}
\end{equation}
where, under the assumption that the dynamics $\phi^\tau$ are
smooth, $x_{_A}$ is determined by the mean value theorem.
Now, letting the graining of phase space become infinitely 
fine (i.e.\ letting $\ell,k \rightarrow \infty$), we obtain:
\begin{equation}
\frac{\Delta_{th} S_I}{\tau} \equiv \lim_{\ell,k \rightarrow \infty} 
\frac{\Delta_{th}S^{GD}_{\ell,k}(X)}{\tau} = 
\int_X \ln {\cal J}(x) ~ \mu(d x) = \sum^{6N}_{j=1} \lambda_j
\label{delta-th}
\end{equation}
where the $\lambda_j$'s are the Lyapunov exponents determined by the 
dynamics $\phi^\tau$.
The sum of the Lyapunov exponents is negative and $\mu$ is singular
with respect to the Lebesgue measure if the system is strictly dissipative
\cite{RU}. Hence, $\Delta_{th}S^{GD}_{\ell,k}(X)/\tau$,
for sufficiently large $\ell$ and $k$, will also be negative in such a case.

Combining this with the assertion that 
$\Delta_eS^{GD}_{\ell,k}(X)$ and $\Delta_{tot}S^{GD}_{\ell,k}(X)$ 
both vanish in the stationary state, allows GD to set:
\begin{equation}
\frac{\Delta_i S_I}{\tau} \equiv \lim_{l,k\rightarrow\infty} 
\frac{\Delta_i S^{GD}_{\ell,k}(X)}{\tau} = -\lim_{l,k\rightarrow\infty} 
\frac{\Delta_{th}S^{GD}_{\ell,k}(X)}{\tau} = - \sum^{6N}_{j=i} \lambda_j \, .
\label{delta-i}
\end{equation}
so that the irreversible entropy production $\Delta_i S^{GD}_{\ell,k}$ 
has the proper positive sign for dissipative systems. The irreversible 
production is here obtained as the ``loss of information'' about the 
probability distribution in going from one level of resolution (that of
${\cal A}_{\ell+1,k}$) to another (that of ${\cal A}_{\ell+2,k-1}$)
level of resolution in the graining of phase space.\footnote{We put in
quotes ``loss of information'' to stress the fact that this does not
directly correspond to the usual (Kolmogorov-Sinai) loss of information
of dynamical systems theory. In fact, the second level of resolution is
not necessarily coarser than the first, it is merely different.
If ${\cal A}$ is a Markov partition, then
${\cal A}_{\ell+2,k-1}$ is coarser than ${\cal A}_{\ell+1,k}$ 
in the stable directions \cite{BobBook}.}
We remark that in the definition of $\Delta_e S^{GD}_{\ell,k}$, 
Eq.(\ref{GDflow}), and hence of
$\Delta_i S^{GD}_{\ell,k}$, the term $S^{GD}_{\ell,k}(\phi^{-\tau}B,t)$
appears. However, there may be no
collection of cells $A$ of ${\cal A}_{\ell,k}$ whose union is 
the set $\phi^{-\tau}B$. 
For this reason the finer partition ${\cal A}_{\ell+1,k}$
had to be introduced in the definition of the GD information
entropy Eq.(\ref{Aprime}) (cf.\ Figure 3).

\subsection{Gaspard's results}
In G's book \cite{PG} a review of his previous work is given,
in which a special kind of partitions was considered: partitions 
whose cells all have the same phase space volume $\varepsilon$. This 
was only for simplicity and does not change the substance of the 
results. Therefore, we will denote G's partition with the symbol 
${\cal A}$.

Gaspard only considers systems of independent points coupled to 
infinite reservoirs (thought to represent driven systems 
of noninteracting particles) and 
proceeds with the construction of a Poisson suspension
measure $\chi$, from which a coarse grained
entropy can be defined. One can see that this coarse grained entropy 
reduces to the GD information entropy plus a rest term (cf.\
Eq.(\ref{POentrop}) below). Gaspard then argues that
this rest term can be made small with respect
to the GD information entropy
by taking the size of the partition cells sufficiently small.
Therefore, the rest term may be neglected, and G's calculations are
then equivalent to GD's calculations.
In particular, the term 
called $\varepsilon$-entropy flow by G, Eq.(8.105) of \cite{PG},
is nothing other than GD's $\Delta_e S^{GD}_{0,1}$, if one starts from a
partition ${\cal A}$ made of equal cells of size $\varepsilon$
and takes ${\cal A}_{\ell,k}$ with $\ell=0, k=1$. 
The same holds for the $\varepsilon$-entropy production, 
Eq.(8.106) of \cite{PG}, which equals GD's $\Delta_i S^{GD}_{0,1}$.
On the other hand, Gaspard only uses dynamics which are phase space 
volume preserving \cite{PG}, and he does not consider the term 
$\Delta_{th} S^{GD}_{\ell,k}$.

Applying G's theory to the case of multibaker dynamics 
one obtains the following relation, Eq.(8.125) of \cite{PG}:
\begin{equation}
\Delta_i S^{GD}_{0,1} = D \frac{(\nabla \rho)^2}{\rho} + \mbox{ 
higher order terms}
\label{PGhigher}
\end{equation}
where $D$ is the diffusion coefficient in the multibaker space $X$, $\rho$
is the density of the points moving through $X$
via baker-dynamics, while $\nabla \rho$ is the corresponding stationary 
state gradient of $\rho$ imposed by the presence of the unequal density 
in the boundary reservoirs. Note that the independence of
the points, which allows the construction of a Poisson suspension,
is crucial here to pass from a description in the phase space 
${\cal M}$ to the ``1-point'' (thought to be 1-particle) space $X$, 
making the operator $\nabla$ a gradient in real space.

The quantity $\Delta_i S^{GD}_{0,1}$ then turns out to have 
the desired form expected from IT for diffusion,
which can be related to the (baker map) diffusion 
coefficient $D$ by, Eq.(8,126) of \cite{PG}:
\begin{equation}
\lim_{\varepsilon\rightarrow 0} 
\lim_{(\nabla \rho/\rho)\rightarrow 0}
\lim_{L\rightarrow\infty} \frac{\rho}{(\nabla \rho)^2} \Delta_i S^{GD}_{0,1}   
= D > 0 ~,
\label{PGdiff}
\end{equation}
where $L$ is the size of the system between the two reservoirs. 
Equations (\ref{PGhigher},\ref{PGdiff}) represent the first 
instance in which IT-like expressions for diffusive 
systems were derived from an area-preserving map.

\subsection{The Breymann-T\'el-Vollmer results}
In the more general multibaker model considered by
BTV, two kinds of coarse grained information entropies were defined:
one, $S_m^{BTV,c}$, using the densities of points in each rectangle
of area $ab$, and another one, $S_m^{BTV,C}$, using the single horizontal
strips (of area $alb$, $asb$ and $arb$ respectively) of each rectangle,
Fig.2. The collection of baker rectangles constitutes
one of the two partitions of the system considerd by BTV
(${\cal A}_{\ell,k}$ in GD's notation), while the collection
of the three horizontal strips of all rectangles constitutes the other 
partition (${\cal A}_{\ell+1,k}$). 
The two coarse grained quantities are
\begin{eqnarray}
&& S_m^{BTV,c} = - a b \varrho_m \log \frac{\varrho_m}{\varrho^*}
\label{BTVCG}\\
&&S_m^{BTV,C} = - a r b \varrho_{m,b} \log \frac{\varrho_{m,b}}{\varrho^*}
- a s b \varrho_{m,c} \log \frac{\varrho_{m,c}}{\varrho^*} -
a l b \varrho_{m,t} \log \frac{\varrho_{m,t}}{\varrho^*} ~,
\label{BTVG}
\end{eqnarray}
where $\varrho^*$ is a constant reference density, $\varrho_{m,b}$ is the 
coarse-grained density on the bottom horizontal strip of cell $m$,
$\varrho_{m,c}$ is the coarse-grained density on the central strip and 
$\varrho_{m,t}$ is the coarse-grained density on the top strip of cell 
$m$. The variation in time of the
coarse grained entropy $S_m^{BTV,c}$ is split into two terms:
the flow term
\begin{equation}
\Delta_e S_m^{BTV,c}(t) = S_m^{BTV,C}(t+\zt) - S_m^{BTV,C}(t) ~,
\label{BTVflow}
\end{equation}
which is assumed to be the same as the total 
variation $S_m^{BTV,C}$;
and the irreversible information entropy
production term
\begin{equation}
\Delta_i S_m^{BTV,c}(t) = \left( S_m^{BTV,c}(t+\zt) - S_m^{BTV,C}(t+\zt)
\right) - \left( S_m^{BTV,c}(t) - S_m^{BTV,C}(t) \right) ~.
\label{BTVirr}
\end{equation}
The sum of $\Delta_e S_m^{BTV,c}$ and $\Delta_i S_m^{BTV,c}$ is
then the total variation of $S_m^{BTV,c}$ in one time step.

The next important ingredient of BTV's approach is the macroscopic limit 
for these multibaker models. This uses an expansion up to second 
order derivatives in terms of the horizontal coordinate $x$ for the 
density:
\begin{equation}
\varrho(x\pm a) = \varrho(x) \pm a \partial_x \varrho(x) + \frac{a^2}{2}
\partial_x^2 \varrho(x) ~.
\label{BTVexpa}
\end{equation}
Furthermore, the system is seen as a biased random walk on the line, so that
one  can attribute to it a given drift velocity $v$ and a given diffusion
coefficient $D$ for each choice of $r$ and $l$. 
The quantities $r,l,v,D$ can then be used to define a
scaling for the parameters $a$ and $\zt$:
\begin{equation}
r = \frac{\zt D}{a^2} \left(1 + \frac{a v}{2 D} \right) ~, \quad
l = \frac{\zt D}{a^2} \left(1 - \frac{a v}{2 D} \right)
\label{BTVscale}
\end{equation}
so that a meaningful fine grained limit is obtained, in which both
$a$ and $\zt$ tend to zero. Then, the following expression results 
for the irreversible information entropy production \cite{BTV}:
\begin{equation}
\zs^{BTV} = \frac{\varrho}{D} \left( v - D \frac {\nabla \varrho}{\varrho}
\right)^2
\label{BTVmaclim}
\end{equation}
which, in the case with $r=l=1/2$, i.e.\ $v=0$, reduces to the
same formula as given by Gaspard, Eq.(\ref{PGhigher}) above, 
except for the higher order terms present in Eq.(\ref{PGhigher}).

Three observations are in order here. 1) The special choice of partitions 
made by BTV is not strictly necessary to obtain the BTV results.
Different choices which are closer to GD's ${\cal A}_{\ell,k}$ are
possible, as explained in the Appendix of \cite{BTV}. 2) It is
possible to keep higher order corrections in the calculations
sketched above, so that terms corresponding to G's higher order
terms of Eq.(\ref{PGhigher}) can also be found within the BTV approach
(cf.\ Ref.\cite{VTB}). 3) The scaling given by Eqs.(\ref{BTVscale}) 
is such that the relaxation time difficulty discussed below in 
Section 4.2 does not appear in the BTV's approach. This scaling has
been recently adopted also by Gaspard and Tasaki in \cite{GT99}.

\section{Difficulties}
The results presented in the previous sections, give rise to a number 
of questions some of which we will discuss here, concentrating on their 
implications for a consistent dynamical theory of IT. In particular we 
will try to identify the range of validity of the results
obtained, pointing
out which problems should in our opinion still be clarified
or overcome.

\subsection{The phase space difficulty}
Relations such as Eqs.(\ref{PGhigher}),(\ref{BTVmaclim}) 
for multibaker maps look similar to those obtained in IT 
where real-space gradients appear. However, this is somewhat 
misleading and due to the simplicity of the map, whose phase space
has in practice only one active dimension, (the direction of the density 
gradient in \cite{PG} or the direction of 
``transport'' in \cite{BTV,GD}), and to the
assumption that the multibaker dynamics are valid
substitutes for independent particle systems. In that case, 
indeed, there can be two situations: a) the system is infinite and can
be described by a Poisson distribution; b) the system is finite, and 
the many-particle distribution factorizes. In both situations we are
allowed to go from a description in the phase space ${\cal M}$
to a description in the 1-point space $X$, where there is only 
one active (real-space) dimension. Then, entropy-like quantities
can only flow in this direction, giving 
necessarily rise to real-space expression such as
Eqs.(\ref{PGhigher}),(\ref{BTVmaclim}).\footnote{Note that if the 
$1$-dimensional chain of baker maps is replaced
by a $d$-dimensional lattice of baker maps, the active dimensions
are $d$, and flows only occur in the $d$-dimensional real-space.} 
This point will be examined further in Subsection 4.4.

Some difficulty emerges when this approach has to be applied to
a wider clss of models than that of multibaker maps. In particular,
interacting particle systems are not compatible with this approach. 
To study these situations
some improvement of the presently developed approach is required.
Indeed, following the general definitions and derivations 
given in \cite{PGpap}--\cite{GD}, one immediately realizes
that in principle the flows and the gradients computed there are
all in terms of {\em phase-space} variables, and not in terms of {\em 
real-space} variables: $\Delta_eS^{GD}_{\ell,k}(B)$ 
represents a flow through the phase space volume $B$. 
Therefore, $\Delta_eS^{GD}_{\ell,k}$ 
given by Eq.(13) could be seen as the substitute for the flow 
${\bf J}_S^G$ (Eq.(8.85) in \cite{PG})
\begin{equation}
{\bf J}_S^G = \left( -\rho \log \rho \right) \dot{x} ~, 
\label{PGcurr}
\end{equation}
which takes place in phase space, in the case that the state of the 
system is represented by a singular measure, for which the Gibbs 
entropy is not defined. In turn,  
${\bf J}_S^G$ is reminiscent of the convective entropy flow of 
IT, cf.\ ${\bf J}_{s,c}$ defined below Eq.(\ref{DEGMloc2}), so that 
$\Delta_eS^{GD}_{\ell,k}(B)$ could be thought of as representing 
${\bf J}_{s,c}$.\footnote{In the case that the space of the system and 
of the reservoirs are combined, as for multibaker models, this flow 
term would account for the total entropy flow.}
However, the IT entropy flow takes place in real space, not in phase 
space, and the phase space cannot be reduced to real space 
if there are interacting particles, or if there is a flow 
in momentum space. Therefore, the diffusion coefficient $D$ present e.g.\
in Eqs.(\ref{PGdiff}),(\ref{BTVmaclim}), in a more general context
would concern diffusion in phase space rather than in real space.
It is not clear, then, how the Gibbs entropy flow
or its coarse grained substitute 
introduced in \cite{PGpap}--\cite{GD}, could be related to
the IT entropy flow. 

\subsection{The relaxation times difficulty}
In IT, the relaxation times of given processes, i.e. their approach from
an initial state to a stationary state, are directly 
related to the transport coefficients. Obvioulsy, speaking of relaxation
times, one should first indicate which physical quantities are observed 
to relax, and which tolerance is accepted in assessing the relaxation. 
In general, when dealing with particle systems, the relaxation time is 
intended to be determined by the relatively short Maxwell relaxation time,
$\zt_M$, which
is typically the time of a few collisions per particle. In fact, the main 
physical observables approximate within measurable errors their 
stationary values in such a time. 
In any case, given the observables in which one is interested 
(e.g.\ smooth functions of phase) and 
the relaxation tolerance, the times needed for these observables 
to approach precisely enough the relevant limiting values are 
determined by the dynamics alone. 
On the contrary, the coarse grained quantities discussed in 
\cite{PG,GD} have relaxation times which strongly depend on 
the size of the cells of the coarse graining partitions. 

This is similar to the problem of portraying the relaxation to 
equilibrium of an Hamiltonian system, by means of a coarse grained 
version of the Gibbs entropy \cite{Tolman}.
However, in our case the situation appears worse. 
Indeed, in equilibrium the Gibbs entropy equals 
the physical entropy of the system, at least. On the 
contrary, in the case of systems evolving towards nonequilibrium stationary 
states, characterized by singular measures, $S_G$ is not even defined in the 
stationary state. Hence, a coarse
grained version of the Gibbs entropy in
the study of relaxation towards nonequilibrium stationary states is
at risk of being even less meaningful than in the case of relaxation towards 
an equilibrium state. We illustrate these facts with a simple example.

Why doesn't the Gibbs entropy
exist in the nonequilibrium stationary states of systems such as 
those described in Section 2? We have already seen that, starting from
an initial state represented by a regular measure, hence with a given 
initial value of the Gibbs entropy, the time evolution is such that
$S_G(t) \rightarrow -\infty$ as $t \rightarrow \infty$. But, we could 
see more in detail why this happens, considering a simplified model of 
a thermostatted system of the kind discussed in section 2.1. The idea
remains valid in general. Let the
initial state be an equilibrium state, described by the 
microcanonical ensemble in a volume of size $1$, and let the 
stationary distribution be confined to a small (not dense)
fractal region of phase space.
This situation corresponds to
a case with high forcing and consequent high dissipation. 
Let us take a phase space partition ${\cal A}_{\ell,k}$,
made of $M$ cells of equal size
$\ze = 1/M$. The corresponding initial coarse
grained information entropy then verifies
\begin{equation}
S^{GD}_{\ell,k}(X,0) =  1 ~.
\label{oneovereps}
\end{equation}
In the following time evolution leading to a nonequilibrium stationary
state, the overall phase space contraction due to dissipation makes the 
probability distribution gradually concentrate on smaller and smaller
regions of phase space,\footnote{Because we assume the attractor
not to be dense in $X$, if the graining is sufficiently fine, there are
cells of the partition which contain parts of the attractor and
other cells which do not.}
until it differs from zero only on a number $L = L(M) < M$ of cells 
of the partition. Assuming for simplicity that the probability to 
find the system in each of these $L$ cells is the same, we then get
\begin{equation}
S_{ss}(M) \equiv \lim_{t \rightarrow \infty} S^{GD}_{\ell,k}(X,t) =
L \left[ - \frac{1}{L} \log\left(\frac{1/L}{1/M}\right)+1 \right] =
\log\left( \frac{L(M)}{M}\right)+1 ~.
\label{Sinfty}
\end{equation}
Therefore, since the fraction $L(M)/M$ tends to zero when $M$ tends to
infinity, we have $S_{ss}(M) \rightarrow -\infty$ for
$(1/M) = \ze \rightarrow 0$.
In other words, the fact that the
phase space probability distribution is rearranged by the time evolution, 
so that sets of zero volume take a probability of $1$ in the stationary 
state, makes the Gibbs entropy diverge to $-\infty$, {\it indicating 
that there is no connection between the Gibbs entropy and the physical 
entropy of the system}. This is still true even if the dissipation is small, 
and all sets of measure $1$ are dense in phase space.

Let us consider then, in more general terms, one thermostatted
system evolving towards a nonequilibrium stationary state, whose 
initial state is represented by a regular measure $\mu_0$, for which 
the Gibbs entropy $S_G(0)$ is defined. In the following evolution, 
$S_G(t)$ gradually diverges to $-\infty$, but at any positive time $t$ 
the distribution $\mu_t$ remains regular, and the corresponding Gibbs
entropy can be approximated better and better by finer and finer
coarse grained entropies, as in Eq.(\ref{SGlim}).
Because of the divergence of $S_G$ and because the size of the partition 
cells needed in the definition of $S^{GD}_{\ell,k}$ can be taken 
arbitrarily small, the total information entropy change, Eq.(\ref{GDtot}), 
can be kept different from zero during arbitrarily long times. Indeed,   
by taking finer and finer partitions,
$\Delta_{tot}  S^{GD}_{\ell,k}(X,t)/\tau$ will approach better and
better, and for longer and longer times, the rate of decrease of the
fine grained information entropy $S_I = S_G/k_B$, given by 
Eq.(\ref{SGrate}), which has a definite negative value of 
order $O(N)$ ($\approx -3 N \langle \za \rangle_{ss}$).
Now, for every fixed $t \ge 0$
(which could exceed $\tau_M$ by any amount),
the state of the system is represented by a probability measure 
$\mu_t$ which has a density $\rho_t$. 
Hence, given any tolerance $\zd > 0$,
and any time increment $\tau > 0$, there will be an $\ze_{\zd,\zt} > 0$ 
such that (cf.\ Fig.4):
\begin{equation}
\left| S^{GD}_{\ell,k}(X,t) - S_G(t)/k_B \right| < \zd \quad \mbox{and } 
~~ \left| S^{GD}_{\ell,k}(X,t+\zt) - S_G(t+\zt)/k_B \right| < \zd ~,
\label{smalldiff}
\end{equation}
if the size of the cells of the partition ${\cal A}_{\ell,k}$ is smaller
than $\ze_{\zd,\zt}$. It follows that
\begin{equation}
\frac{\Delta_{tot}  S^{GD}_{\ell,k}(X,t)}{\tau} = \frac{S_G(t+\zt) -
S_G(t)}{k_B\tau} + O\left(\frac{\zd}{\zt}\right) =
\frac{1}{k_B} \frac{d S_G}{d t} + O(\zt) + O\left(\frac{\zd}{\tau}\right) 
= O(N) ~,
\label{OrdN}
\end{equation}
instead of 
$\Delta_{tot} S^{GD}_{\ell,k}(X,t)/\tau \approx 0$, 
since we can take that $O(\zd/\zt)$ is $O(1)$ or less, because $\zt$ is
fixed a priori.

Therefore, unphysical, partition dependent, relaxation times have been
introduced through the coarse-graining procedure, which are
extraneous to the dynamics of the system.

The approach of BTV \cite{BTV} seems to avoid the problem of the
relaxation times, because of the way its authors defined their
{\em macroscopic limit}, cf.\ subsection 3.3. In this approach one 
does not take finer and finer partitions of each baker map rectangle;
one only increases the number of rectangles between the reservoirs 
reducing their side $a$ and the length of the time step
$\tau$, in such a way that Eqs.(\ref{BTVscale}) are verified.
Then, the fact that the number of time steps $n$ has to increase in 
order for the entropy to reach its stationary value could be
balanced by the decrease of $\tau$, so that $n \tau$ may converge
to a finite value. From this point of view, then, the BTV
macroscopic limit should be preferred.

\subsection{The difficulty of unphysical definitions}
The relaxation times problem points out further difficulties: the very
definition of the entropy flow and irreversible entropy production could 
be flawed. Indeed, the total 
rate of variation of the {\em real} IT entropy, $\Delta_{tot} S$, relaxes to 
its stationary value zero in the relatively short time $\tau_M$, implying 
that this rate of variation becomes (and remains) smaller than a small
$\zd$ within a time of order $O(\zt_M)$. Therefore, for any arbitrarily 
chosen time $t$ larger than $\tau_M$, if the cells of the partition are 
smaller than $\ze_{\zd,\zt}$, Eq.(\ref{OrdN}) yields:
\begin{equation}
k_B \frac{\Delta_{tot} S^{GD}_{\ell,k}(X,t)}{\zt} -
\frac{\Delta_{tot} S(t)}{\zt} = k_B O(N) ~,
\label{ratediff}
\end{equation}
where the second term on the l.h.s. is $O(\zd)$ or less, and the second 
is of order $k_B O(N)$. Assuming with GD that 
$\Delta_e S^{GD}_{\ell,k}(X,t) = 0$, one can rewrite Eq.(\ref{ratediff}),
with (\ref{GDbala}) and (\ref{deGMbala}), in the form 
\begin{equation}
\frac{k_B}{\tau} \left[ \Delta_{th} S^{GD}_{\ell,k}(X,t) + 
\Delta_i S^{GD}_{\ell,k}(X,t) \right] - 
\frac{1}{\tau} \left[
\Delta_e S(t) + \Delta_i S(t) \right]  = k_B O(N) ~.
\label{absurd}
\end{equation}
In particular, consider a value $t$ which is not necessarily exceedingly 
large, but larger than $\zt_M$. Without taking an extremely fine 
partition, and recalling that then $k_B O(N)$ is approximately equal
to $-\Delta_i S/\zt$, as seen in subsection 2.1, we can write
\begin{equation}
\frac{k_B}{\tau} \Delta_{th} S^{GD}_{\ell,k}(X,t) \approx
\frac{1}{\tau} \Delta_e S(t) 
-\frac{k_B}{\tau} \Delta_i S^{GD}_{\ell,k}(X,t) ~.
\label{absurd1}
\end{equation}
But this contradicts IT, according to which the quantity on the
left hand side of Eq.(\ref{absurd1}), being the overall coarse grained
entropy flow, should approximately equal
only the first term on the right hand side.
Therefore, at least 
either $\Delta_i S^{GD}_{\ell,k}$ or $\Delta_{th} S^{GD}_{\ell,k}$
cannot be correct.
For, if $ \Delta_i S^{GD}_{\ell,k}$
is of order $O(N)$, as the irreversible entropy production should be,
then $\Delta_{th} S^{GD}_{\ell,k}$ is not the external
entropy flow, while it should be,
and if $\Delta_i S^{GD}_{\ell,k}$ is not of order $O(N)$, then it cannot be 
an irreversible entropy production. 

Therefore, the agreement between $ \Delta_i S^{GD}_{\ell,k}$
and $ \Delta_{th} S^{GD}_{\ell,k}$ in the stationary state with their IT
counterparts achieved only after a coarse graining dependent relaxation 
time, appears accidental.
This suggests that the very definitions of the
various terms on the right hand side of Eq.(\ref{GDbala}) cannot
be physically correct. Again, the macroscopic limit of BTV may fix
this problem.

\subsection{The multibaker space difficulty}
Simple dynamical systems such as the multibaker maps
are very useful in understanding many aspects of chaotic dynamics.
In a sense we could say that they play in this context a role
similar to that of exactly soluble models in equilibrium
statistical mechanics. However, the solvability often goes at 
the expense of the degenerately simple nature of the models themselves, 
which, in the case of multibaker chains, becomes cause of concern when 
one wants to identify certain features of the multibaker dynamics with 
known IT properties of real systems. 

In particular, in order to speak of a quantity in some way related
to the Gibbs entropy, one would need a phase space,
in which (at least up to canonical transformations)
half of the dimensions represent the ``configurations'' 
of the system in space and the other half represents the ``momenta''.
Then, in that phase space a coarse grained information entropy can be
defined, which multiplied by $k_B$ and in the limit of fine graining
becomes the Gibbs entropy itself (if it exists).
To do this in a multibaker chain one has
to identify position and momentum variables. These are
assumed by BTV to be represented by the horizontal (along the chain)
direction, and by the vertical (along the thickness of the chain)
direction, respectively. 
If this identification is correctly carried out, then the multibaker
phase space, being $2$-dimensional, could only be a substitute for
a $1$-particle model in one dimension. Alternatively, the
$1$-particle distribution could be used to describe a gas of
identical noninteracting particles, perhaps in the presence 
of obstacles, as in the Lorentz gas. However, even the picture
of the gas of independent particles is at odds with the BTV 
interpretation of the multibaker dynamics: points at different heights 
along the vertical direction of the baker rectangles
can move in exactly the same way under the baker dynamics of the model, 
while points which are at the same height can move in totally different 
directions. Therefore, the vertical direction has nothing to do with 
momentum space.

This problem
could perhaps be fixed by interpreting the baker dynamics and phase
space differently. Like in \cite{PGpap,PG}, one could assume that the
multibaker phase space mimics a Poincar\'e section of a particle
system such as the Lorentz gas. The dynamics is followed from rectangle
to rectangle like a moving particle in the Lorentz gas
is followed from collision to collision. In that case, the problem
of identifying the momentum variables is not so important anymore. 
However, two new problems emerge. In the first place, the coarse grained
entropy of the system should be expressed in terms of all the phase
space variables, and not just in terms of the variables of the
Poincar\'e section. 
Therefore, the contributions due to the
direction of the flow need to be worked out.
However, perhaps more importantly, with this interpretation one 
would also lose 
the possibility of taking the BTV macroscopic limit, because one 
cannot assume that particle collisions occur at a rate which is 
coarse graining dependent. This because in G's interpretation one
time step is the time elapsed between two collisions, but the time
step goes to zero as the graining is made finer and finer in the BTV 
macroscopic limit.

\subsection{Factorizability, entropy and entropy production}
We now try to understand under which conditions the results 
obtained for multibaker models in the 1-point space are valid for
independent many-points systems. This leads us also to an analysis of
the relationship between the definitions
of entropy and entropy production rate, given in \cite{PGpap}-\cite{GD}.
We follow \cite{PGpap,PG}, also in order to point 
out some subtleties of that approach.

Consider a distribution $\mu$ in the 1-point baker space.
The associated Poisson suspension, corresponding
to a ``gas'' of infinitely many independent   
points \cite{PG,CFS}, is characterized by the 
probability measure
\begin{equation}
\chi(C_{B,N}) = \frac{\mu(B)^N}{N!} e^{-\mu(B)} ~; \quad \mbox{with } ~
\chi(C_{B,N} \cap C_{B',N'}) = \chi(C_{B,N}) \chi(C_{B',N'})
\quad \mbox{if }~ B \cap B' = \emptyset ~,
\label{Poissonsusp}
\end{equation}
where $\chi(C_{B,N})$ is the probability of finding $N$ points
in the 1-point volume $B$, and $C_{B,N}$ is the corresponding set in 
the phase space of the Poisson suspension. It is in this phase space
that G can define his coarse grained $\varepsilon$-entropy for a boundary 
driven system. If then $\{ B_i \}$ is a partition of the 1-point space 
in cells of volume $\varepsilon$, this entropy takes the form 
(cf. Eqs.(8.98),(8.99) of \cite{PG}):
\begin{eqnarray}
S_\varepsilon (\{ B_i \}) &=& - \sum_i \sum_{N=0}^\infty \chi(C_{B_i,N})
\log \chi(C_{B_i,N}) \\
&=& \sum_i \mu(B_i) \log \frac{e}{\mu(B_i)} + {\cal R}(\varepsilon)
\label{POentrop}
\end{eqnarray}
where the rest term is
\begin{equation}
{\cal R}(\varepsilon) = - \sum_i e^{-\mu(B_i)} \sum_{N=0}^\infty
\frac{\mu(B_i)^N}{N!} \log N!      ~.
\label{RofEps}
\end{equation}
The terms inside the external sum are of order $O(\mu(B_i)^2)$ and
higher, hence can be neglected in Eq.(\ref{POentrop}) if
$\mu(B_i)$ is small. This is obtained for a multibaker system of
length $L$, by taking sufficiently fine partitions, i.e.\ $B_i$
of sufficiently small volume. 

This step is of fundamental importance for the IT-like results 
of \cite{PGpap,PG} to be valid for a many-point system. In fact,
these results are obtained using the first term of 
Eq.(\ref{POentrop}) only, neglecting ${\cal R}(\varepsilon)$. 
However, in principle, ${\cal R}(\varepsilon)$ could be large,
compared to the first term in Eq.(\ref{POentrop}), because
the macroscopic limit ($L \rightarrow \infty$ for multibaker maps) 
has to be taken before the fine graining limit (cf.\ discussion 
below Eq.(8.126) of \cite{PG}).
This means that, however small the volume of $B_i$ might be,
as long as it does not vanish, its measure $\mu(B_i)$ will be 
large in general, since the density will be large, making 
${\cal R}(\varepsilon)$ also large. 
In this case, neglecting ${\cal R}(\varepsilon)$ 
implies that the IT-like results are not derived from the many-points 
distribution $\chi$, but instead from a kind of information entropy 
defined through the 1-point distribution $\mu$.
Now, because $\mu$ is not normalized, it cannot be a factor of 
a many-points distribution, making those results valid
only for a 1-point system.\footnote{To have a normalized 1-point 
distribution, one would have to implement the boundary conditions 
in a different way, using, e.g., a compact phase space with bulk 
dissipative dynamics, like in thermostatted systems, or with 
biased dynamics in certain regions (e.g.\ the walls)
of the system \cite{HHP}.}

Gaspard overcomes this difficulty in an ingeneous way,
by splitting the fine graining limit into two steps:
The limit of vanishing linear size of the partition cells 
along the unstable direction is taken {\it before} the macroscopic limit,
the remaining limit of vanishing linear size of the cells $B_i$ 
along the stable direction is taken {\it after} the macroscopic limit,
so that the macroscopic limit ``interrupts'' the fine graining 
process. This way, the volumes of the partition cells are made vanishing 
before the macroscopic limit is taken, ${\cal R}(\varepsilon)$ 
can be neglected and the results presented in Section 3.\ follow.

However, the fine graining limit, in particular the part taken 
before the macroscopic limit, comes at the 
cost of losing the coarse grained information entropy 
(which then diverges). Similar considerations hold also for the 
results of BTV and GD, therefore it seems that the knowledge of 
both the entropy and its production rate cannot be given together
with the present approaches, as already noted for the 
thermostatted approach.

\subsection{The thermostat difficulty}
The need for the term $\Delta_{th}S^{GD}_{\ell,k}(X)$ in Eq.(\ref{GDtot})
was deduced by GD from the fact that the information 
entropy flow $\Delta_eS^{GD}_{\ell,k}(X)$ cannot represent an entropy flow 
between the system and its environment, if the system is closed or 
periodic. Nevertheless, this seems to be at odds with IT.
Indeed, $\za(x){\bf p}_i$ is merely introduced to 
enable the externally driven {\em dynamical} system to reach a stationary 
state as is done automatically by a thermostat in $\Delta_e S$ in IT.
In fact, it is for that reason that $\za(x){\bf p}_i$ is usually referred 
to as the 
``thermostat'' of the system. However, dynamically, this term 
has nothing to do with a real thermostat and, in fact, 
it appears in Eqs.(\ref{eqsmot}) as a Lagrange multiplier, due 
to the application to the $N$-particle system of Gauss' (purely 
dynamical) principle of minimum constraint, to make the system
preserve its kinetic or total energy in the course of time. 
In the derivation of Eqs.(\ref{eqsmot}), no use is
made of the properties of any other dynamical system constituting 
a thermostat. Therefore,
an interpretation of $\za(x){\bf p}_i$ as representing an actual 
physical thermostat, which absorbs the dissipative energy created in 
the system by the external forces ${\bf{F}}^e$, and has an explicit
representation in the entropy balance Eq.(\ref{GDtot}), is an 
interpretation which appears to have no basis in the purely 
dynamical nature of the equations (\ref{eqsmot}) themselves.

What can be said, instead, is that Eqs.(\ref{eqsmot}) 
serve as a convenient tool to describe a system in a nonequilibrium
state by purely dynamical means, without incurring the technical
difficulties posed by infinitely large reservoirs. 
That a {\bf real} system would not settle on a  nonequilibrium
state without the presence of a {\bf real} thermostat, seems 
irrelevant in the analysis of the dynamics of Eqs.(\ref{eqsmot}). 

\section{Discussion}
{\bf 1. } The above discussion on the coarse grained approach to 
a complete dynamical
theory of IT pointed out difficulties which we found in the
current formulations. 
Therefore it seems to us that a coarse grained entropy approach based
on $S_G$ does not provide
at present a satisfactory connection with IT. The same can be said about 
thermostatted systems. However, for the latter systems the irreversible 
entropy production is at least unambiguously known at any time: in the 
transient as well as in the stationary state.
This is not the case in the coarse grained description. 
Indeed, we pointed out various difficulties which affect 
the treatments of IT provided by BTV, G and GD. The approach of
BTV could avoid the problems connected with the transient states,
and it is worthwhile to further study this topic,
but the phase space dynamics seems to be very special. On the other hand,
the approach of G and GD was intended to
describe stationary states only \cite{GDG}, despite the full 
time dependent treatment they give \cite{PGpap,PG,GD}.

\vskip 10pt \noindent

\vskip 10pt \noindent
{\bf 2. } It seems that the possibility of identifying in thermostatted
systems other contributions, beyond the irreversible entropy
production term, occurring in IT, is not obvious.
On the basis of our analysis we would argue that, so far, the dynamics
of thermostatted systems allows us only to identify the
irreversible entropy production rate. It is obvious that a stationary
state of a {\bf real} system, with a given irreversible 
entropy production rate will be affected by an equal and
opposite divergence of an entropy flow.
Nevertheless, this does not emerge from the
dynamics of thermostatted systems.
The connection of dynamical properties of thermostatted systems
with the term div${\bf J}_{S,tot}$ occurring in IT, remains 
therefore unclear.

\vskip 10pt \noindent
{\bf 3.}
Although the idea of a possible connection between
coarse-graining, information loss and entropy changes discussed here 
is very intriguing, as far as we can tell,
it does not seem to work in its present form for macroscopic systems,
as long as one connects it with $S_G$, which diverges to $-\infty$.
The fact that the rate of change $-\dot{S}_G$  
equals the irreversible entropy production
rate of thermostatted systems does not seem to be a  reason sufficient 
to assume that $S_G$ itself has any direct connection with 
the entropy of such a system.
Morever, the connection of the information
loss used here with the usual (Kolmogorov-Sinai) information loss, 
if any,
and its relevance for the calculation of the IT entropy is also
not clear to us.
Therefore, it seems to us that
further study of the connection of the dynamics of
particle systems in nonequilibrium states and IT is still required.

\vskip 25pt

\section*{Acknowledgements} 
\vskip 10pt 
\noindent
We would like to thank F.\ Bonetto,
C.P.\ Dettmann, J.R.\ Dorfman, P.\ Gaspard, T.\ Gilbert, T.\ T\'el
and J.\ Vollmer for inspiring discussions.
LR gratefully acknowledges support from GNFM-CNR (Italy) and
from MURST (Italy).
EGDC gratefully acknowledges support from the US Department of Energy
under grant DE-FG02-88-ER13847.

\newpage
\section*{Figure captions}
{\bf Figure 1.} One time step in the evolution of the infinite
multibaker chain. The squares with labels $0,1,...,L$ constitute
the system. The others constitute the reservoirs.
One time step corresponds
to one application of $\phi$, which moves the poins with a given shade
to the points with the same shade. This time evolution
is volume preserving, hence it does not affect the density of points. 
Starting from any initial distribution, we see how the densities
of the baths enter into the system. In the stationary state, only
the blackest and the white densities fill the cells of the system.

\vskip 20pt
\noindent
{\bf Figure 2.} One time step of the evolution of the BTV multibaker
model. Unlike in Fig.1, there is phase space
contraction here if $l \ne r$. The same dynamics of Fig.1 is obtained if
$s=0$ and $l=r=1/2$. One time step moves the points of the different
vertical strips of rectangle $m$
with a given shade to rectangle $m-1$, $m$ or $m+1$,
respectively, with the same shade.

\vskip 20pt
\noindent
{\bf Figure 3.} From left to right we have the partition ${\cal A}$, in
in the baker square $m$, the partition $\phi^{-1}{\cal A}$, and the 
partition $\phi^{-1}{\cal A} \vee {\cal A}$, in the baker square
$m+1$, respectively. The preimage $\phi^{-1} A$ of
every set $A$ in cell $m$, which is
the union of cells of ${\cal A}$, can be partitoned by cells of 
$\phi^{-1}{\cal A} \vee {\cal A}$, while it cannot be partitioned
by cells of ${\cal A}$ itself.

\vskip 20pt
\noindent
{\bf Figure 4.} In the left panel is depicted the decay of the IT entropy 
from a transient to a stationary state, which takes
a time of the order of the Maxwell relaxation time (or is
determined by an appropriate transport coefficient). Here we have
assumed that the initial (equilibrium) entropy is higher than the
steady state entropy.
The right panel shows the decay of the coarse grained entropies for various 
partition sizes (curves labelled by $(1)$, $(2)$ and $(3)$), and the 
divergence of the Gibbs entropy (thickest line). All the coarse grained 
entropies start from the same value, and eventually settle on 
a plateau.
However, they remain close (within a
distance $\zd$, say) to $S_G$ for longer and longer times if the relevant 
partitions are finer and finer. Curve $(1)$ corresponds to the coarsest
partition.
The region delimited by curve $(2)$, by $S_G$ and by the two vertical
(solid line) segments is made of points whose distance from $S_G$ is less
than $\zd$.

\newpage

\pagestyle{empty}

\begin{figure}
\centering\epsfig{figure=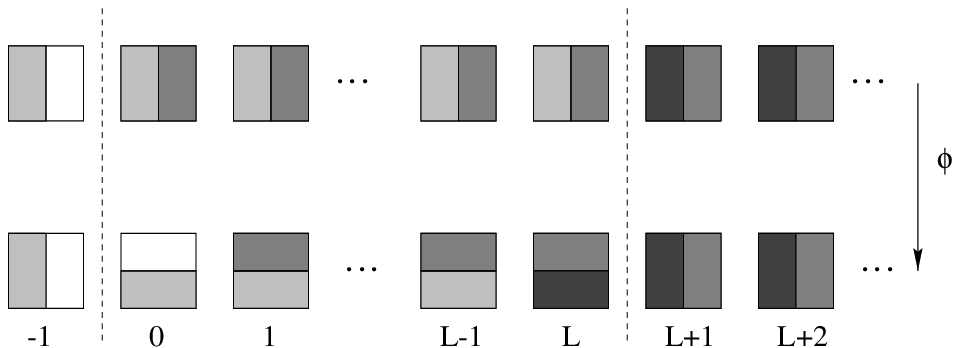,width=10cm}
\caption{\baselineskip=12pt \small }
\label{gaspard}
\end{figure}

\newpage

\begin{figure}
\centering\epsfig{figure=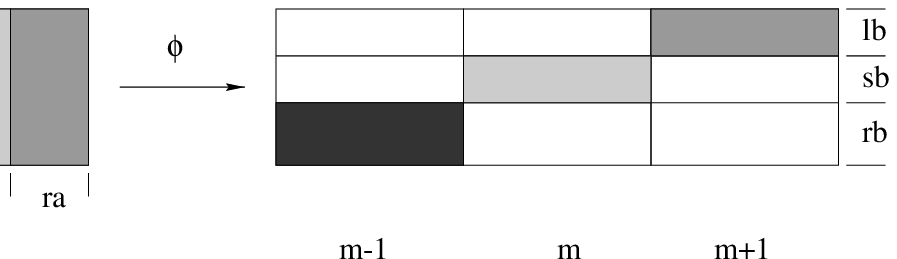,width=10cm}
\caption{\baselineskip=12pt \small }
\vskip 4cm
\label{vollmer}
\end{figure}
\newpage

\begin{figure}
\centering\epsfig{figure=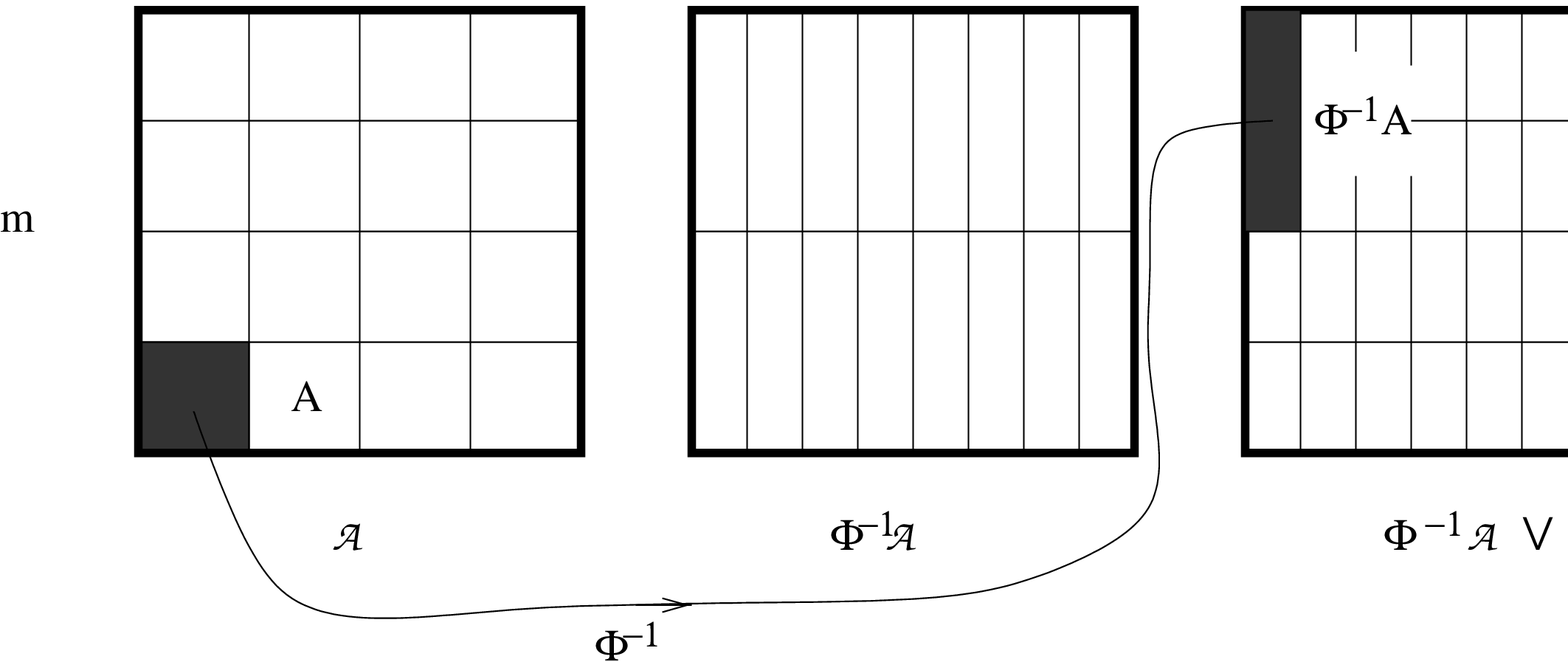,width=10cm}
\caption{\baselineskip=12pt \small }
\label{partit2}
\end{figure}

\newpage
\begin{figure}
\centering\epsfig{figure=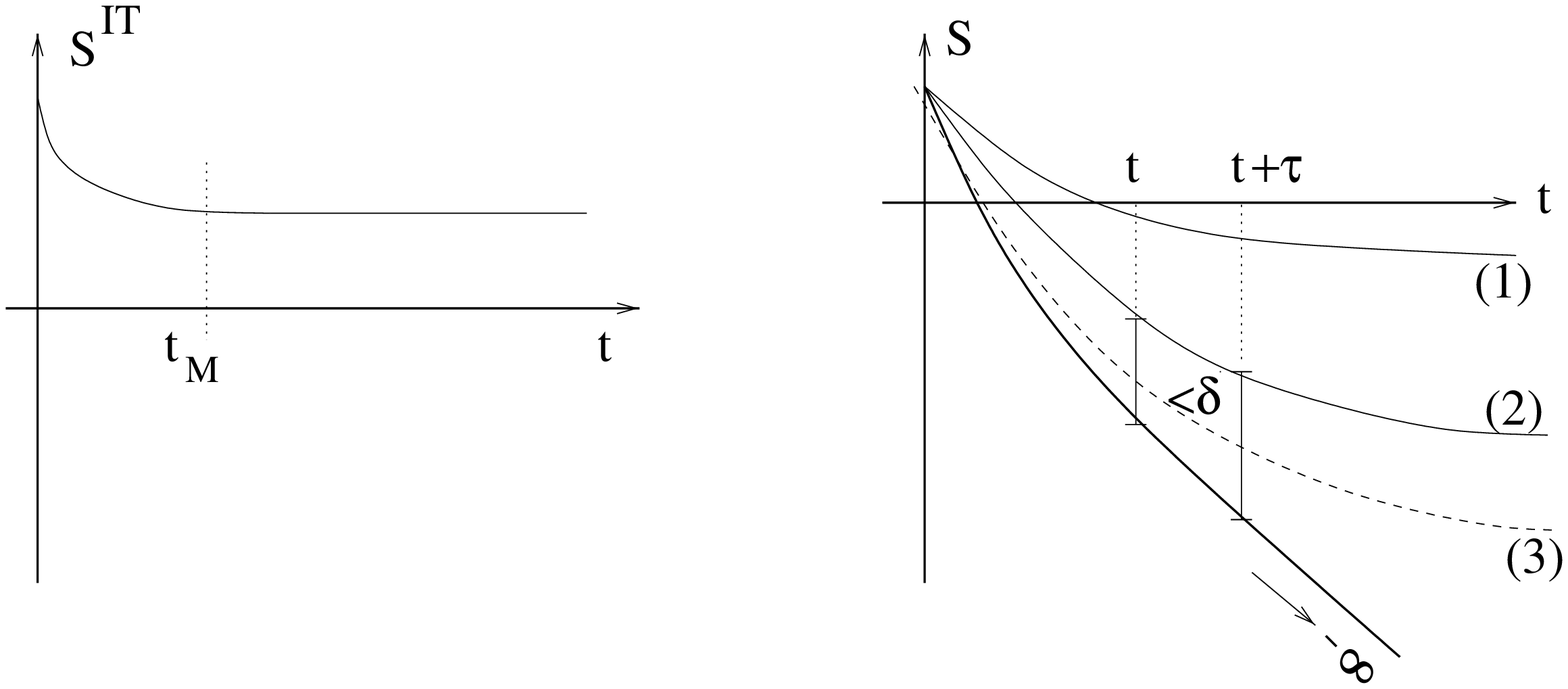,width=10cm}
\caption{\baselineskip=12pt \small }
\label{entrocu}
\end{figure}

\end{document}